\documentclass[a4paper,UKenglish,cleveref,autoref,thm-restate,pdfa]{lipics-v2021}

\pdfoutput=1

\title{Beyond Rule-Based Mutation Testing: Test-Aware Mutant Generation Using Large Language Models}
\titlerunning{Test-Aware Mutant Generation Using LLMs}
\category{Technical Track Paper}

\author{Nils Kiele}{Department of Electrical and Software Engineering, University of Calgary, AB, Canada}{nils.kiele@ucalgary.ca}{https://orcid.org/0009-0000-9049-9324}{}
\author{Zainab Saad}{Department of Electrical and Software Engineering, University of Calgary, AB, Canada}{}{https://orcid.org/0009-0007-0702-6827}{}
\author{Zirui Wang}{Department of Electrical and Software Engineering, University of Calgary, AB, Canada}{}{https://orcid.org/0009-0002-2366-1269}{}
\author{Steve Drew}{Department of Electrical and Software Engineering, University of Calgary, AB, Canada}{}{https://orcid.org/0000-0003-4527-2635}{}
\author{Samira {Ebrahimi Kahou}}{Department of Electrical and Software Engineering, University of Calgary, AB, Canada}{}{https://orcid.org/0009-0005-6283-3811}{}

\authorrunning{N. Kiele, Z. Saad, Z. Wang, S. Drew, and S. Ebrahimi Kahou}
\Copyright{Nils Kiele, Zainab Saad, Zirui Wang, Steve Drew, and Samira Ebrahimi Kahou}



\supplementdetails[swhid={swh:1:dir:753cf1771d9178c5f28653eb3b401645689508de;origin=https://github.com/codingWhale13/test-aware-mutants;visit=swh:1:snp:d5503902c906af374c271693613f0cd931b4867b;anchor=swh:1:rev:fa0313ea93d5e1bcadc2576b486ed0ae178186ea}]{Software}{https://github.com/codingWhale13/test-aware-mutants}
\supplementdetails{Dataset}{https://zenodo.org/records/20195774}

\acknowledgements{We thank Prof. Ahmad Abdel Latif for helpful discussions and feedback.}

\funding{The Machine Intelligence and Reasoning Lab at the University of Calgary is funded by NSERC, CIFAR, DRAC, Alberta Innovates, Schulich Momentum, and Denvr Dataworks.}

\nolinenumbers

\usepackage{booktabs}
\usepackage{tcolorbox}
\tcbuselibrary{breakable}
\tcbuselibrary{skins}
\usepackage{xspace}
\usepackage{glossaries-prefix}

\glsdisablehyper
\newacronym[prefixfirst={a\ },prefix={an\ }]{llm}{LLM}{large language model}
\newacronym{ci}{CI}{continuous integration}
\newacronym{iqr}{IQR}{interquartile range}
\newacronym{rag}{RAG}{retrieval-augmented generation}
\newacronym{sota}{SOTA}{state-of-the-art}
\newcommand{\eg}{e.g.\@\xspace}
\newcommand{\ie}{i.e.\@\xspace}
\newcommand{\vs}{vs.\@\xspace}
\newcommand{\quotes}[1]{``#1''}

\definecolor{setupblue}{HTML}{C3E5F6}
\definecolor{setupgreen}{HTML}{A1D4CB}
\definecolor{setupred}{HTML}{E5B2BA}
\definecolor{tagcolor}{HTML}{888888}
\definecolor{syscolor}{HTML}{2E7D32}
\definecolor{placeholdercolor}{HTML}{C62828}
\definecolor{taonly}{HTML}{FFF3CD} 
\definecolor{taonlyborder}{HTML}{E0C97A}
\definecolor{success}{HTML}{4CAF50}
\definecolor{undetected}{HTML}{2196F3}
\definecolor{rejected}{HTML}{FF9800}
\definecolor{invalid}{HTML}{F44336}
\newcommand{\invalidLabel}{\textcolor{invalid}{\textsc{Invalid}}\@\xspace}
\newcommand{\rejectedLabel}{\textcolor{rejected}{\textsc{Rejected}}\@\xspace}
\newcommand{\successLabel}{\textcolor{success}{\textsc{Success}}\@\xspace}
\newcommand{\undetectedLabel}{\textcolor{undetected}{\textsc{Undetected}}\@\xspace}
\newcommand{\xmltag}[1]{\textcolor{tagcolor}{\textlangle #1\textrangle}}
\newcommand{\tahighlight}[1]{\colorbox{taonly}{#1}}

\EventEditors{Robert Feldt, Maria Paasivaara, Daniel Mendez, Stefan Wagner, and Marvin Mu\~{n}oz Bar\'{o}n}
\EventNoEds{5}
\EventLongTitle{20th International Symposium on Empirical Software Engineering and Measurement (ESEM 2026)}
\EventShortTitle{ESEM 2026}
\EventAcronym{ESEM}
\EventYear{2026}
\EventDate{October 8--9, 2026}
\EventLocation{Munich, Germany}
\EventLogo{}
\SeriesVolume{394}
\ArticleNo{49}

\makeatletter
\newcommand\abstractsubparagraph{\@startsection{subparagraph}{5}{\z@}%
                                       {-1ex \@plus-1ex \@minus -.2ex}%
                                       {-1.5ex \@plus 2ex}
                                      {\color{dagpubGray}\sffamily\normalsize\bfseries}}
\makeatother

\begin{document}

\maketitle

\begin{abstract}
\abstractsubparagraph{Background.} Mutation testing evaluates test-suite adequacy by injecting synthetic faults into program code. However, traditional rule-based tools often generate large numbers of trivial, redundant, or equivalent mutants that limit their practical use for identifying gaps in a test suite. While recent \gls{llm}-based approaches generate more realistic faults, most remain test-blind: The model sees only the source code and cannot reason about what existing tests already cover. Ignoring such tests means neglecting context that could help generate higher-quality mutants and thus stronger tests to patch the remaining test suite gaps.
\abstractsubparagraph{Aims.} We propose \emph{test-aware} mutant generation, in which \pgls{llm} receives the problem statement, canonical solution and base tests in a single prompt, and must generate a nontrivial mutant that passes the base unit tests.
\abstractsubparagraph{Method.} We evaluate this approach across a set of five \glspl{llm} -- Gemini 3.1 Pro, Gemini 3 Flash, GPT 5.1 Codex Mini, GPT 4.1 Mini, Qwen3-32B -- on the HumanEval and MBPP benchmarks. The extended EvalPlus test suites serve as an automated oracle to verify whether surviving mutants represent genuine bugs.
\abstractsubparagraph{Results.} Test-aware prompting yields verified fault rates of 87.7\% (HumanEval) and 79.1\% (MBPP), meaning these mutants pass all base tests but are caught by the oracle. This vastly outperforms the matched test-blind prompting (which yields only 12.2\% and 23.0\%, respectively) and the traditional rule-based tool mutmut (4.4\% and 5.7\%). While fault subtlety (the fraction of extended tests a mutant fails) remains comparable across all three methods, test-awareness minimizes the computational cost per verified fault, compared to test-blind prompting.
\abstractsubparagraph{Conclusions.} Exposing \pgls{llm} to existing unit tests shifts mutant generation from untargeted bug injection toward effective discovery of weaknesses in an existing test suite. Our work establishes a concrete foundation for future research to scale test-aware mutant generation to production-level environments.
\end{abstract}

\glsresetall

\section{Introduction}

Mutation testing is a well-established technique for evaluating and improving the quality of a software test suite by injecting synthetic faults, called \emph{mutants}, into source code and checking whether existing tests detect them~\cite{jia2010analysis, papadakis2019mutation}. A surviving mutant is one that passes all tests and reveals a gap in the test coverage~\cite{just2014mutants}. Despite the benefits of mutation testing, its practical adoption remains limited. A key challenge is the equivalent mutant problem: traditional rule-based frameworks like the Python tool mutmut\footnote{\url{https://mutmut.readthedocs.io/en/latest/}} generate many mutants that are syntactically different but semantically identical to the original code. Distinguishing these from genuine faults is generally undecidable~\cite{papadakis2019mutation}. Empirical studies estimate that 5--20\% of generated mutants are equivalent~\cite{madeyski2013overcoming}. In industrial settings, the computational cost of executing tests against large mutant pools and the manual effort needed to triage them are the main bottlenecks~\cite{petrovic2018industrial}. For example, calculating the mutation score (ratio of \quotes{killed} mutants to total mutant count) on the two billion lines of Google code is generally infeasible~\cite{petrovic2018state}. Furthermore, even non-equivalent mutants often provide little value, as many are trivially killed by basic assertions rather than exposing concealed logical flaws. This motivates our search for fewer, higher-value mutants that reveal genuine test-suite weaknesses.

\Glspl{llm} are a promising tool in efficiently generating mutants. Recent work has shown that \gls{llm}-generated mutants are more diverse and behaviorally closer to real bugs than those from rule-based tools, having a 1.8$\times$ higher real fault detection rate~\cite{wang2026comprehensive}. However, existing \gls{llm}-based approaches such as LLMorpheus~\cite{tip2025llmorpheus} and BugFarm~\cite{ibrahimzada2025challenging} use \emph{test-blind} prompting where the model is provided only the source code and has no knowledge of what the existing tests already cover. Without this context, the \gls{llm} cannot explicitly target gaps in test coverage and may instead produce mutants that are redundant, trivially caught, or semantically equivalent. Instead of improving test quality, these low-quality mutants present unnecessary additional cost either in terms of human review or computational resources required for repeated execution at every future code deployment. In contrast, a \emph{test-aware} model can reason about the behavioral constraints already covered by the tests and focus its mutations on what the suite misses. Although these techniques enhance mutation testing, they require iterative prompting and are computationally expensive. To reduce this overhead, we propose a single-prompt, test-aware mutant generation approach.

More concretely, the \gls{llm} is provided with both the source code and the existing test suite in one prompt and asked to generate a subtle logical bug that survives the current tests. This approach isolates the impact of test-awareness in prompting \glspl{llm} for mutant generation and investigates whether the reasoning of the model alone (without a feedback loop) is enough to efficiently generate mutants. However, there is a methodological challenge: verifying whether a surviving mutant is a genuine, non-obvious fault, rather than an equivalent program, typically requires manual review. To address this, we use the Python-based EvalPlus benchmark~\cite{liu2023evalplus}, which extends HumanEval~\cite{chen2021humaneval} and MBPP~\cite{austin2021mbpp} with 80$\times$ and 35$\times$ more tests, respectively. The base test suites (7.7 tests on average per HumanEval problem and 3 for MBPP) serve as the developer-written tests the \gls{llm} must evade. The extended suites act as an automated oracle, classifying a mutant that passes the base tests but fails at least one extended test as a verified, non-equivalent fault. 

Through extensive experiments with five \glspl{llm} -- Gemini 3.1 Pro, Gemini 3 Flash, GPT 5.1 Codex Mini, GPT 4.1 Mini, Qwen3-32B -- we study various aspects of test-aware prompting for mutant generation and provide a comparison to its matched test-blind and exhaustive rule-based alternatives. Crucially, we show that including the existing test cases in the \gls{llm}'s context yields an increase of up to 7$\times$ in the rate of verified fault generations. We then measure how many oracle test cases are triggered to evaluate the subtlety of generated mutants; a subtle mutant fails on fewer oracle tests. Finally, our cost analysis demonstrates that while rule-based techniques remain the cheapest overall, test-aware prompting significantly reduces the amortized token cost per verified fault compared to the test-blind baseline.

\section{Related Work}

Traditional mutation testing tools rely on basic syntactic mutation operations such as altering true/false values or modifying numeric literals. One such tool is the Python library mutmut. Although simple and fast, it produces many low-value candidates that do not compile, are trivially caught by basic tests or are semantically equivalent to the initial code. This motivates our work to focus on mutant quality over quantity by generating fewer yet more informative mutants. Our intuition is that a test-aware \gls{llm} can reason about the semantic blind spots in the code to produce targeted non-trivial bugs for important edge cases.

\subparagraph*{Test-Blind LLM Mutation.}
Several approaches use \glspl{llm} to generate mutants from source code without access to existing unit tests. $\mu$BERT~\cite{degiovanni2022mu} generates mutants using the pretrained CodeBERT~\cite{feng2020codebert} model to predict masked code replacements. LEAM~\cite{tian2022learning} trains a syntax-guided encoder-decoder on real-world bug-fix patches from GitHub for syntactically correct mutant generation. These methods produce mutants that resemble real faults more closely than semantics-agnostic operators, but they are limited in syntactic accuracy and mutant diversity. LLMorpheus~\cite{tip2025llmorpheus} prompts an \gls{llm} to fill placeholders in JavaScript code, producing mutants that are more realistic than those from the \gls{sota} rule-based tool StrykerJS\footnote{\url{https://github.com/stryker-mutator/stryker-js}}. BugFarm~\cite{ibrahimzada2025challenging} uses attention analysis to find the least attended locations in the code and generate hard-to-detect bugs. The main limitation of these approaches is that they are test-blind and do not condition mutant generation on what the current tests already check. LLMut~\cite{wang2026comprehensive} is closest to our work and considers varying levels of contextual information in \pgls{llm} prompt. The authors report substantially higher real-bug detection than rule-based mutation, but with a lower compilation rate and higher duplication. Interestingly, while one of their prompts includes existing unit tests, it performs slightly worse than the variant that omits them. We hypothesize that this performance drop occurs because LLMut only appends the tests as passive context. In contrast, our approach succeeds by explicitly structuring the prompt as an adversarial task, instructing the model to bypass the provided tests to produce high-quality, targeted mutants.

\subparagraph*{Adversarial and Iterative LLM Approaches.}
AdverTest~\cite{chang2026test} improves the fault detection on the Defects4J dataset~\cite{just2014defects4j} by using two \gls{llm} agents in an adversarial loop where one generates mutants to target blind spots, and the other generates tests. 
MutGen~\cite{wang2026mutation} is an iterative mutation-guided \gls{llm}-based test generation framework that incorporates mutant feedback (whether it survived or was killed by the tests) into the prompt for improved unit tests. Nafis et al. developed an iterative test generation pipeline where mutant score and test coverage are used as feedback signal for test case refinement~\cite{nafis2025llm}. These approaches show that feedback-driven workflows can be effective, but they also require repeated compilation, execution and re-prompting. A related iterative approach on the test-generation side lets the \gls{llm} form hypotheses about how to kill specific mutants and refine tests using execution feedback~\cite{straubinger2025mutation}. Compared to these iterative approaches, our proposed test-awareness yields strong results using only a single interaction with the \gls{llm}, reducing cost.

\subparagraph*{Mutation-Guided and Coverage-Guided Test Generation.}
A complementary line of work uses mutation or coverage information to improve \gls{llm}-generated tests rather than to generate mutants. MuTAP~\cite{dakhel2024effective} uses rule-based mutants to iteratively improve \gls{llm}-generated tests. Meta's TestGen-LLM provides both source code and existing tests to the model when improving unit tests~\cite{alshahwan2024automated}. CoverUp~\cite{altmayer2025coverup} iteratively targets uncovered code in Python regression test generation, and SymPrompt~\cite{ryan2024code} exploits the multi-step reasoning capability of \glspl{llm} and generates test cases in multiple reasoning stages guided by execution path analysis. These approaches do not generate mutants, but they are consistent with our central idea that giving \glspl{llm} richer test-related context improves generation quality.

\subparagraph*{LLMs for Equivalent Mutant Detection.}
Equivalent mutants remain a fundamental challenge for scaling mutation testing. An empirical study on using \glspl{llm} for equivalent mutant detection evaluates several strategies such as fine-tuned code embeddings and prompting strategies on Java mutant pairs~\cite{tian2024large}. The results show that \gls{llm}-based techniques outperform prior equivalent mutant detection methods by an average F1 score of 35.69\%. This work uses \glspl{llm} to filter equivalent mutants after generation, while we aim to reduce low-value mutant generation by giving the model access to the tests.

To the best of our knowledge, no prior work empirically studies the direct impact of test-awareness on \gls{llm}-based mutant generation. Test-blind methods cannot directly find test-suite gaps without iterative methods requiring expensive multi-step feedback. Coverage- and mutation-guided test-generation approaches show that test-related context helps \gls{llm} reasoning, but they apply that insight to the test side rather than mutant generation.

\section{Method and Experimental Setup}

\begin{figure}[t]
\centering
\includegraphics[width=\linewidth]{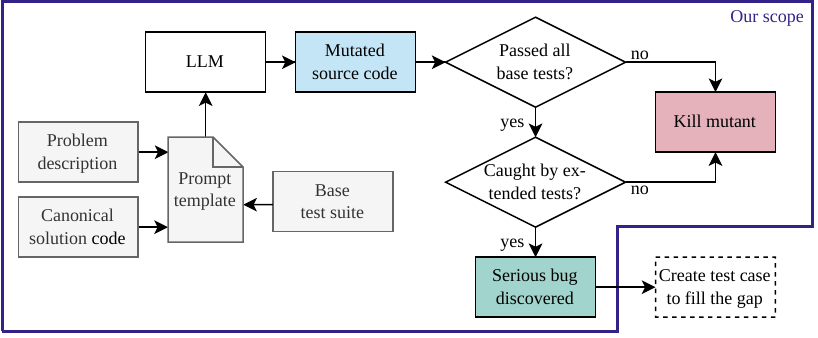}
\caption{Our proposed test-aware mutation pipeline. The \gls{llm} generates a \sethlcolor{setupblue}\hl{candidate mutant}, which is filtered by the base tests and then validated using the extended EvalPlus oracle, to automatically determine whether the mutant is a \sethlcolor{setupgreen}\hl{serious gap in the test suite} or \sethlcolor{setupred}\hl{not}.}
\label{fig:experimental-setup}
\end{figure}

Our objective is to generate mutants that reveal weaknesses in an existing test suite rather than producing arbitrary bugs. We formulate mutant generation as an adversarial task. Given a problem statement $P$, canonical solution code $C$, and base tests $T_\text{base}$, \pgls{llm} must produce a mutant $C'$ that preserves the original function signature, passes $T_\text{base}$, and fails at least one test on the hidden oracle $T_\text{oracle}$. For every problem, model, and prompt configuration, the \gls{llm} generates exactly one mutant without compilation feedback, test-execution feedback, or iterative repair (except in the iterative ablation, see \cref{sec:iterative-execution-results}). Our single-shot test-aware mutation pipeline is described below and illustrated in \cref{fig:experimental-setup}. Mutants are assigned an outcome label, as described in \cref{tab:outcomes}.
\begin{enumerate}
\item \textbf{Prompting.} An \gls{llm} receives the problem statement, canonical solution and -- in the test-aware setting -- the base tests. The prompt asks for a subtle logical bug that is executable and passes the provided base tests.
\item \textbf{Extraction.} The raw response is parsed to extract the mutant code and discard the surrounding conversational text or markup. If the mutant omits the essential imports, the canonical solution imports are used instead to compile the mutant.
\item \textbf{Base-Test Filtering.} The extracted mutant is executed against the base tests and discarded if it throws an error, times out, or fails. This ensures that only eligible mutants are passed to the oracle, which (1) saves cost in running oracle tests on fewer mutants, (2) provides a controlled setting to compare test-awareness vs. test-blindness, and (3) verifies that the mutant is not a trivial bug, but rather has potential to be a genuine fault.
\item \textbf{Oracle Evaluation.} Surviving mutants are executed against the EvalPlus extended tests. A mutant that fails at least one oracle test is classified as a verified fault.
\end{enumerate}

\begin{table}[thb]
\caption{Each \gls{llm}-generated mutant is assigned one of the four mutually exclusive outcome classes, the meaning of which is explained in this table.}
\centering
\begin{tabular}{ll}
\toprule
\textbf{Mutant Type} & \textbf{Definition} \\
\midrule
\invalidLabel & Mutant does not compile or times out during base-test execution. \\
\rejectedLabel & Mutant runs without error, but is killed by at least one base test. \\
\successLabel & Mutant evades the base tests but is detected by the oracle. \\
\undetectedLabel & Mutant survives both the base tests and the oracle. \\
\bottomrule
\end{tabular}
\label{tab:outcomes}
\end{table}

\subparagraph*{Benchmarks and Oracle Design.}
We evaluate generated mutants on EvalPlus~\cite{liu2023evalplus}, which extends two Python code-generation benchmarks with substantially stronger test suites for 164 HumanEval problems~\cite{chen2021humaneval} and 378 MBPP problems~\cite{austin2021mbpp}. While EvalPlus was designed to benchmark code generation quality, we repurpose it to determine the quality of the \gls{llm}-generated mutant. We view the few base tests for each problem as high-quality but incomplete developer-written tests that a good mutant must evade, while the much larger extended test suite acts as an automated oracle. This design lets us automatically determine the distinction between mutants that are arbitrary syntax modifications \vs those that represent real semantic faults, overlooked by the base tests.
We selected EvalPlus over traditional mutation testing benchmarks, such as Defects4J~\cite{just2014defects4j}, because it enables the automated protocol described above to reliably classify mutants at scale without manual review. Furthermore, the authors of the original EvalPlus study~\cite{liu2023evalplus} empirically confirm that the base tests in HumanEval and MBPP are often insufficient. By exposing flaws in code solutions that would previously have been accepted, their study provides direct evidence that the extended tests are significantly stronger. While these extended test suites are not perfect either, they serve as a useful tool for evaluating our approach.

\subparagraph*{Models and Baselines.}
We evaluate five \glspl{llm}: Gemini 3.1 Pro, Gemini 3 Flash, GPT 5.1 Codex Mini, GPT 4.1 Mini, and Qwen3-32B. These model choices represent a mix of open-source (Qwen3-32B), frontier \gls{sota} models (Gemini 3.1 Pro) and more computationally efficient versions (Gemini 3 Flash, GPT 5.1 Codex Mini, GPT 4.1 Mini). These specific models have been used in recent software engineering literature for tasks such as vibe-coding benchmark~\cite{tran2026vibe}, software issue resolving~\cite{tao2026swe}, code refactoring~\cite{thillen2026codetaste}, and multi-requirement programming tasks~\cite{rontogiannis2026interactive}.
All \gls{llm}-based experiments use temperature 0 to evaluate the \glspl{llm} in deterministic setting; however, this does not guarantee exact reproducibility because \glspl{llm} are non-deterministic by nature and produce stochastic outputs. We compare test-aware prompting against matched test-blind prompting and mutmut.

\subparagraph*{Model Prompts.}
We use two types of prompts, with two variants each. The design rationale for each variant is described below and summarized in \cref{tab:prompt-families}. The prompt families are dubbed \textsc{agent1} (\textbf{A}dversarial \textbf{G}eneration of \textbf{E}vasive \textbf{N}on-equivalent \textbf{T}ransformations~1) and \textsc{agent2}. Each has a test-aware (\textsc{ta}) and a matched test-blind (\textsc{tb}) variant that differ only in whether the base tests are included in the context. These prompt variants allow for a clean ablation of test-awareness. 
All prompts (see \cref{tab:prompt-families} below) share the same task: to generate a single mutant containing a subtle logical bug, given the problem statement and canonical solution. We study two effects: the contribution of test context (\textsc{ta} \vs \textsc{tb}) and the impact of reasoning structure (\textsc{agent1}'s free-form style \vs \textsc{agent2}'s template). 

\begin{table}[thb]
\caption{Summary of prompt variants used in our study. \textsc{ta} = test-aware; \textsc{tb} = test-blind.}
\label{tab:prompt-families}
\centering
\begin{tabular}{ccc}
\toprule
\textbf{Prompt} & \textbf{Tests Given?} & \textbf{Reasoning Scaffold} \\
\midrule
\textsc{agent1-ta} & Yes & Free-form \\
\textsc{agent1-tb} & No& Free-form \\
\textsc{agent2-ta} & Yes & Structured template \\
\textsc{agent2-tb} & No& Structured template \\
\bottomrule
\end{tabular}
\end{table}

We now describe the prompts \textsc{agent1} and \textsc{agent2}, followed by their full system and user section definitions. Content that appears \emph{only} in the \textsc{ta} variant is marked with a \textcolor{syscolor}{\sethlcolor{taonly}\hl{yellow background}} in the full prompt. Placeholders shown in \textcolor{placeholdercolor}{red} are filled at runtime with the problem statement, canonical solution, and base tests.

\enlargethispage{1.5\baselineskip}
\textsc{AGENT1} is the primary prompt. The test-aware variant \textsc{agent1-ta} has the base tests along with the source code and instructs the model to (1) analyze the tests for blind spots on valid inputs, (2) plan a minimal mutation, and (3) \quotes{mentally dry-run} the mutation against each base test. These reasoning steps are elicited by the \gls{llm} inside \texttt{<thinking>} tags. The test-blind variant \textsc{agent1-tb} uses the same system prompt and output format but omits the base tests from the user message.

\begin{tcolorbox}[
title={\textbf{AGENT1} -- Test-Aware (\textbf{TA}) and Test-Blind (\textbf{TB}) Variants},
colback=white, colframe=black!70, coltitle=white, colbacktitle=black!70,
fonttitle=\small\bfseries, breakable, sharp corners,
boxrule=0.5pt, left=4pt, right=4pt, top=4pt, bottom=4pt,
before skip=20pt, after skip=20pt
]
\small
\textbf{SYSTEM:}\\[2pt]
\textcolor{syscolor}{
You are an adversarial software tester and quality-assurance agent. Your goal is to introduce a subtle, realistic logic bug (a \quotes{mutant}) into the provided Python function.}

\begin{tcolorbox}[
colback=taonly, colframe=taonlyborder,
sharp corners, boxrule=0.4pt,
left=4pt, right=4pt, top=2pt, bottom=2pt,
before skip=4pt, after skip=4pt
]
\textcolor{syscolor}{
Your objective is to exploit a semantic blind spot in the Base Tests. The mutated code MUST pass all the provided Base Tests, but it should fail on broader edge cases or uncommon inputs.}
\end{tcolorbox}

\textcolor{syscolor}{Constraints:}\\
\textcolor{syscolor}{1. Preserve the original function signature and all required imports.}\\
\textcolor{syscolor}{2. The modified code must compile successfully without syntax errors.}\\
\textcolor{syscolor}{3. Make the bug subtle and realistic. Do not completely rewrite the function or introduce obvious crashes.}

\begin{tcolorbox}[
colback=taonly, colframe=taonlyborder,
sharp corners, boxrule=0.4pt,
left=4pt, right=4pt, top=2pt, bottom=2pt,
before skip=2pt, after skip=2pt
]
\textcolor{syscolor}{4. The mutated code MUST pass EVERY provided Base Test.}
\end{tcolorbox}

\textcolor{syscolor}{5. The bug MUST manifest on VALID inputs that adhere to the problem description.}\\
\textcolor{syscolor}{6. Do not introduce arbitrary \quotes{magic numbers}, epsilon values, or floating-point precision hacks.}

\textcolor{syscolor}{Format your output EXACTLY as follows:}\\[2pt]
\xmltag{thinking}\\
\textcolor{syscolor}{1. \tahighlight{Analyze the Base Tests to }identify a blind spot on VALID inputs.}\\
\textcolor{syscolor}{2. Plan a minimal mutation to exploit this.}\\
\textcolor{syscolor}{3. Perform a mental dry-run \tahighlight{against the Base Tests} to ensure it passes.}\\
\xmltag{/thinking}\\[2pt]
\xmltag{mutant}\\
\textcolor{syscolor}{[Insert fully modified Python code here, with no markdown formatting]}\\
\xmltag{/mutant}

\tcblower
\textbf{USER:}\\[2pt]
Problem statement:\\[4pt]
\textcolor{placeholdercolor}{\{problem\_statement\}}

\begin{tcolorbox}[
colback=taonly, colframe=taonlyborder,
sharp corners, boxrule=0.4pt,
left=4pt, right=4pt, top=2pt, bottom=2pt,
before skip=2pt, after skip=2pt
]
Base Tests:\\
\textcolor{placeholdercolor}{\{base\_tests\}}
\end{tcolorbox}
Canonical Source Code:\\
\textcolor{placeholdercolor}{\{canonical\_solution\}}\\[4pt]
Generate the mutant.
\end{tcolorbox}

\textsc{AGENT2} describes the \gls{llm} reasoning process in a more structured way. The test-aware variant \textsc{agent2-ta} requires the \gls{llm} to fill the fields in a fixed template: (1) a short summary of the function's computation, (2) base-test coverage (input types, value ranges, edge cases), (3) a list of two to three identified blind spots, (4) a short mutation plan specifying the exact line and change type, and (5) a per-test trace confirming correctness. This prompt is designed to reduce reasoning shortcuts and force \gls{llm} to explicitly analyze gaps in test coverage. Its test-blind counterpart \textsc{agent2-tb} only has the function summary and mutation plan fields, because coverage analysis is not possible without tests.

\begin{tcolorbox}[
title={\textbf{AGENT2} -- Test-Aware (\textbf{TA}) and Test-Blind (\textbf{TB}) Variants},
colback=white, colframe=black!70, coltitle=white, colbacktitle=black!70,
fonttitle=\small\bfseries, breakable, sharp corners,
boxrule=0.5pt, left=4pt, right=4pt, top=4pt, bottom=4pt,
before skip=20pt, after skip=20pt
] 
\small
\textbf{SYSTEM:}\\[2pt]
\textcolor{syscolor}{
You are an adversarial mutation tester. Introduce exactly one subtle logic bug into the given Python function. \tahighlight{The mutant MUST pass every base test but fail on other valid inputs.}}

\textcolor{syscolor}{Reason inside \xmltag{thinking} tags using this exact structure, keep each field to one or two sentences maximum:}\\[2pt]
\xmltag{thinking}\\
\textcolor{syscolor}{WHAT IT DOES: [One sentence: the function's core computation and return value]}

\begin{tcolorbox}[
colback=taonly, colframe=taonlyborder,
sharp corners, boxrule=0.4pt,
left=4pt, right=4pt, top=2pt, bottom=2pt,
before skip=4pt, after skip=4pt
]
\textcolor{syscolor}{BASE TEST COVERAGE:}\\
\textcolor{syscolor}{\quad - Input types tested: [e.g.\ positive ints, non-empty lists]}\\
\textcolor{syscolor}{\quad - Value ranges tested: [e.g.\ n in [1..10], lists of length 2--5]}\\
\textcolor{syscolor}{\quad - Edge cases tested: [e.g.\ duplicates, empty input, single element]}\\[2pt]
\textcolor{syscolor}{BLIND SPOTS: [List 2--3 valid input scenarios the base tests do NOT exercise.]}
\end{tcolorbox}

\textcolor{syscolor}{MUTATION PLAN: [One sentence: exactly what line to change and how.]}

\begin{tcolorbox}[
colback=taonly, colframe=taonlyborder,
sharp corners, boxrule=0.4pt,
left=4pt, right=4pt, top=2pt, bottom=2pt,
before skip=2pt, after skip=2pt
]
\textcolor{syscolor}{BASE TEST TRACE: [For each base test input, confirm the mutant output matches expected]}
\end{tcolorbox}

\xmltag{/thinking}\\[4pt]
\textcolor{syscolor}{Then output the complete modified function:}\\
\xmltag{mutant}\\
\textcolor{syscolor}{[full function, no markdown]}\\
\xmltag{/mutant}

\tcblower 
\textbf{USER:}\\[2pt]
Problem statement:\\
\textcolor{placeholdercolor}{\{problem\_statement\}}\\[4pt]
Original code:\\
\textcolor{placeholdercolor}{\{canonical\_solution\}}

\begin{tcolorbox}[
colback=taonly, colframe=taonlyborder,
sharp corners, boxrule=0.4pt,
left=4pt, right=4pt, top=2pt, bottom=2pt,
before skip=2pt, after skip=2pt
]
Base tests to bypass:\\
\textcolor{placeholdercolor}{\{base\_tests\}}
\end{tcolorbox}
Generate the mutant. 
\end{tcolorbox}

\subparagraph*{Evaluation Metrics.}
For the quantitative analysis of the efficacy of different prompting approaches, we use \quotes{effectiveness} in terms of success rate, \quotes{quality} in terms of fault subtlety, and \quotes{token cost} for generating a mutant. These metrics are calculated as follows:

\begin{alphaenumerate}
\item \textbf{Effectiveness:} We measure the \emph{success rate} as the fraction of generated mutants classified as \successLabel: \[M_{\text{eff}} = \frac{|\{\text{mutants classified as \successLabel}\}|}{|\{\text{all generated mutants}\}|}.\] A higher $M_{\text{eff}}$ means that the \gls{llm} generates more verified, non-equivalent faults.
\enlargethispage{\baselineskip}
\item \textbf{Quality:} We measure fault subtlety using the \emph{oracle kill rate}, defined for each \successLabel mutant $m$ as the fraction of extended oracle tests that detect it: \[M_{\text{sub}}(m) = \frac{|\{t \in T_{\text{oracle}} \mid t \text{ fails on } m\}|}{|T_{\text{oracle}}|}.\] A lower $M_{\text{sub}}$ indicates a narrower fault that is triggered by fewer oracle tests and is therefore harder to detect with a randomly chosen assertion. In addition to the mean, we report the distribution of $M_{\text{sub}}$ across all \successLabel mutants, providing rich information about whether a method produces a long tail of very subtle faults.
\item \textbf{Token Cost:} We report two cost measures, the \emph{raw output tokens} generated by the \gls{llm} and \emph{amortized cost per verified fault}. The amortized cost is calculated by normalizing the sum of input, thinking and output tokens by the number of \successLabel mutants: \[M_{\text{cost}} = \frac{\text{total tokens (input + output + thinking)}}{|\{\text{mutants classified as \successLabel}\}|}.\] A configuration that uses more tokens per prompt but achieves a higher success rate can still have a lower $M_{\text{cost}}$ than a prompt that wastes most of its generations on \invalidLabel or \rejectedLabel outcomes. For the rule-based baseline mutmut, we do not report token cost, since its computational expense lies in execution time rather than API usage; the comparison with mutmut is therefore limited to $M_{\text{eff}}$ and $M_{\text{sub}}(m)$.
\end{alphaenumerate}

\section{Results}

\begin{figure}[thb]
\centering
\includegraphics[width=\linewidth]{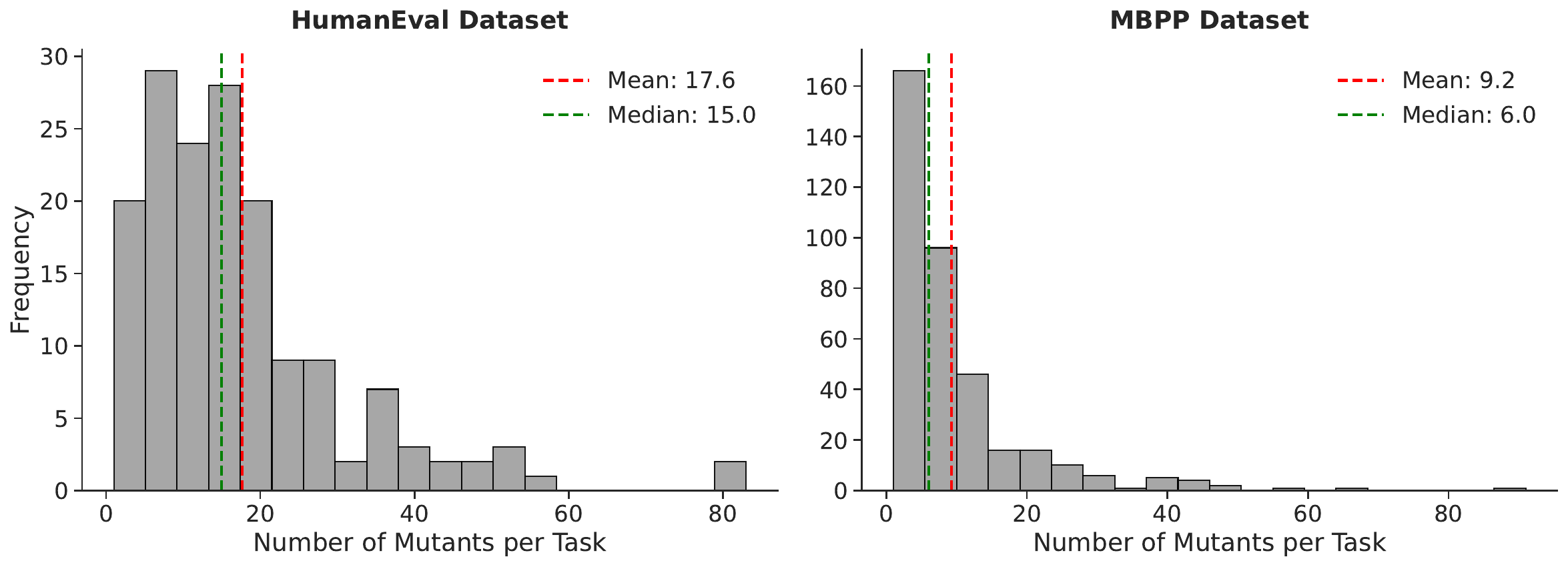}
\caption{Mutant counts per task, generated by mutmut. We use the default settings to run mutmut, resulting in varying numbers of mutants per task, depending on the canonical source code.}
\label{fig:mutmut-counts-per-task}
\end{figure}

To maximize reproducibility, we evaluate all \glspl{llm} with temperature 0. Because this makes the \gls{llm} mostly deterministic, we only generate a single mutant per model-prompt-task combination. In contrast, mutmut's exhaustive rule-based generation produces a significantly larger mutant count: around 18 mutants per HumanEval task and 9 mutants per MBPP task on average (see \cref{fig:mutmut-counts-per-task} for details). Thus, we present the mutmut results separately in \cref{tab:status-overview-mutmut}, showing that the vast majority ends up \invalidLabel or \rejectedLabel.

\begin{figure}[thb]
\centering
\includegraphics[width=\linewidth]{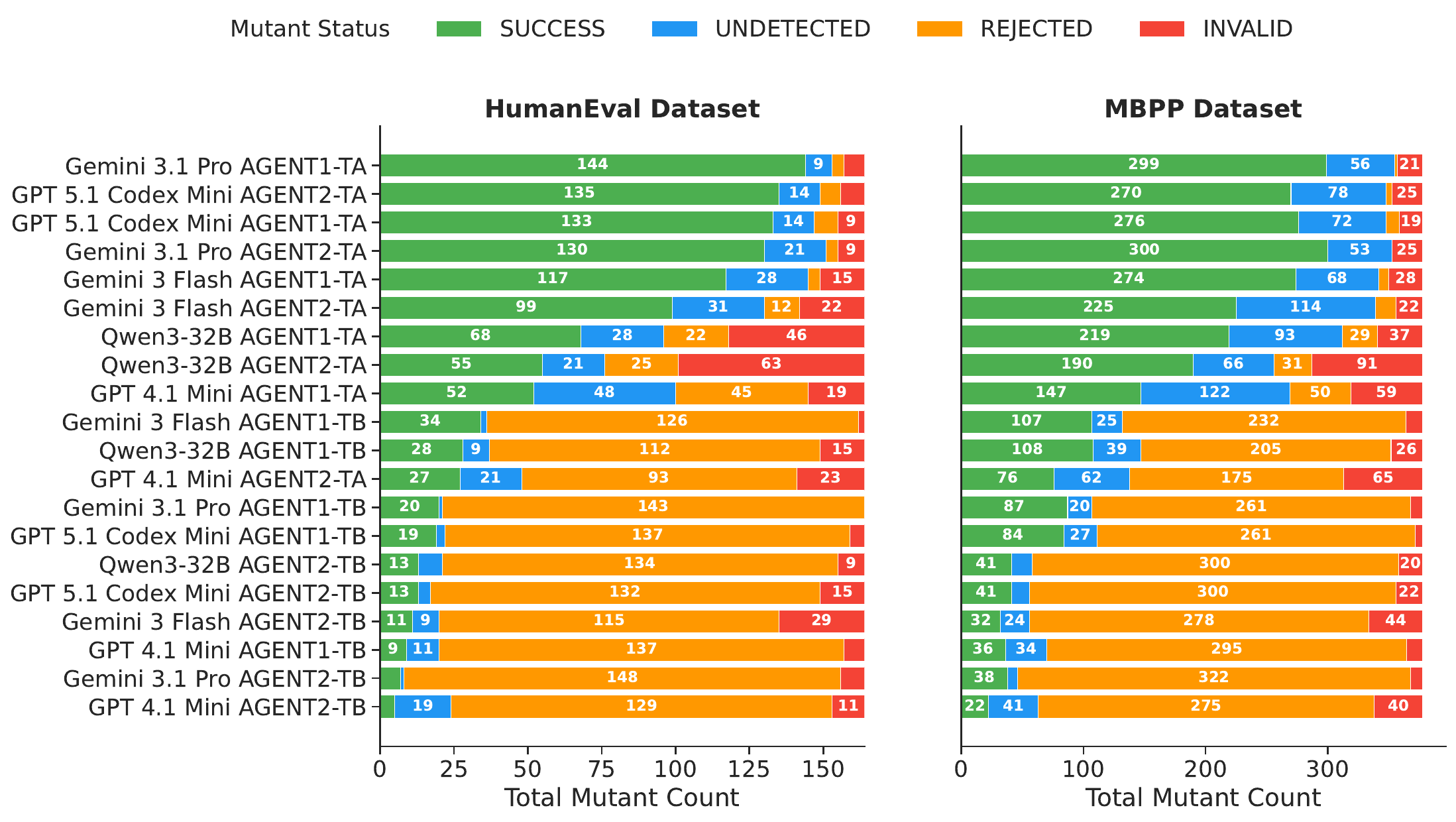}
\caption{Distribution of mutant classifications across the HumanEval and MBPP datasets.}
\label{fig:status-overview}
\end{figure}

The \gls{llm}-based results in \cref{fig:status-overview} reveal that, as we move from test-blind to test-aware prompting, there is a shift in mutant outcomes from \rejectedLabel to \successLabel. For example, \textsc{agent1-tb} generates predominantly \rejectedLabel mutants: 68\% (112) -- 87\% (143) are executable mutants failing on the base tests for HumanEval and 54\% (205) -- 78\% (295) for MBPP across all \glspl{llm}. The proportion of \rejectedLabel and \successLabel changes drastically for the test-aware \textsc{agent1-ta}. For 164 HumanEval problems, the best-performing model, Gemini 3.1 Pro, generates 144 \successLabel mutants (87.8\% \successLabel rate) and only 4 \rejectedLabel. Except for GPT 4.1 Mini, every test-aware model outperforms all other models (even smaller ones) in the test-blind scenario in terms of success rate. The ranking of model-prompt combinations regarding success rate is similar for MBPP, establishing a clear trend. This confirms that without visibility of the base tests, \glspl{llm} introduce changes that existing assertions already cover. We provide the full results in \cref{tab:main-results}.

\begin{table}[thb]
\caption{Status distribution of mutants generated by mutmut. Note that the default setting used for mutmut generates multiple mutants per task, making the total mutant count larger than the number of tasks (2841/164 for HumanEval, 3427/378 for MBPP).}
\centering
\begin{tabular}{llllll}
\toprule
Dataset & \successLabel & \undetectedLabel & \rejectedLabel & \invalidLabel & Total \\
\midrule
HumanEval & 126 & 179 & 1473 & 1063 & 2841 \\
MBPP & 195 & 197 & 1519 & 1516 & 3427\\
\bottomrule
\end{tabular}
\label{tab:status-overview-mutmut}
\end{table}

\begin{table*}[thb]
\caption{Evaluation of \gls{llm}-based mutant generation on 164 HumanEval and 378 MBPP problems under different prompting strategies. For each dataset and model-prompt configuration, we report counts of \invalidLabel (Inv), \rejectedLabel (Rej), \successLabel (Succ), and \undetectedLabel (Und) mutations.}

\label{tab:main-results}
\centering
\scriptsize
\setlength{\tabcolsep}{2.3pt}
\renewcommand{\arraystretch}{1.1}

\resizebox{\textwidth}{!}{
\begin{tabular}{llccccccccccccccccccccc}
\toprule

& & &
\multicolumn{4}{c}{\textsc{agent1-ta}} &
\multicolumn{4}{c}{\textsc{agent2-ta}} &
\multicolumn{4}{c}{\textsc{agent1-tb}} &
\multicolumn{4}{c}{\textsc{agent2-tb}} \\

\cmidrule(lr){4-7}
\cmidrule(lr){8-11}
\cmidrule(lr){12-15}
\cmidrule(lr){16-19}

\textbf{Dataset} & \textbf{Tool} & \textbf{Total}
& Inv & Rej & Succ & Und
& Inv & Rej & Succ & Und
& Inv & Rej & Succ & Und
& Inv & Rej & Succ & Und \\

\midrule

\multirow{5}{*}{\textbf{HumanEval}}
& gemini-3-flash & 164 
& 15 & 4 & 117 & 28
& 22 & 12 & 99 & 31 
& 2 & 126 & 34 & 2 
& 29 & 115 & 11 & 9 \\

& gemini-3.1-pro & 164
& \textbf{7} & \textbf{4} & \textbf{\underline{144}} & \textbf{9}
& 9 & 4 & 130 & 21
& 0 & 143 & 20 & 1
& 8 & 148 & 7 & 1 \\

& gpt-4.1-mini & 164
& 19 & 45 & 52 & 48
& 23 & 93 & 27 & 21
& 7 & 137 & 9 & 11
& 11 & 129 & 5 & 19 \\

& gpt-5.1-codex-mini & 164
& 9 & 8 & 133 & 14
& \textbf{8} & \textbf{7} & \textbf{\underline{135}} & \textbf{14}
& 5 & 137 & 19 & 3
& 15 & 132 & 13 & 4 \\

& qwen3-32b & 164
& 46 & 22 & 68 & 28 
& 63 & 25 & 55 & 21 
& 15 & 112 & 28 & 9 
& 9 & 134 & 13 & 8 \\

\midrule

\multirow{5}{*}{\textbf{MBPP}}
& gemini-3-flash & 378
& 28 & 8 & 274 & 68
& 22 & 17 & 225 & 114
& 14 & 232 & 107 & 25
& 44 & 278 & 32 & 24 \\

& gemini-3.1-pro & 378
& \textbf{21} & \textbf{2} & \textbf{\underline{299}} & \textbf{56}
& \textbf{25} & \textbf{0} & \textbf{\underline{300}} & \textbf{53}
& 10 & 261 & 87 & 20
& 10 & 322 & 38 & 8 \\

& gpt-4.1-mini & 378
& 59 & 50 & 147 & 122
& 65 & 175 & 76 & 62
& 13 & 295 & 36 & 34
& 40 & 275 & 22 & 41 \\

& gpt-5.1-codex-mini & 378
& 19 & 11 & 276 & 72
& 25 & 5 & 270 & 78
& 6 & 261 & 84 & 27
& 22 & 300 & 41 & 15 \\

& qwen3-32b & 378
& 37 & 29 & 219 & 93 
& 91 & 31 & 190 & 66 
& 26 & 205 & 108 & 39 
& 20 & 300 & 41 & 17 \\

\bottomrule
\end{tabular}
}

\end{table*}

\enlargethispage{-1\baselineskip}
We also observe that test-aware prompting produces more \undetectedLabel mutants than test-blind (\eg, 9 \vs 1 on HumanEval with \textsc{agent1} for Gemini 3.1 Pro). Given the larger number of \rejectedLabel mutants in the test-blind case, this occurs because test-blind models generate many trivial faults caught by the base tests, leaving fewer to reach the oracle. Also, the number of \invalidLabel mutants depends mostly on the \gls{llm} used, with smaller models (Qwen3-32B, Gemini 3 Flash, GPT 4.1 Mini) producing more such errors than larger models (GPT 5.1 Codex Mini, Gemini 3.1 Pro). This is likely due to the smaller models' limited capacity to understand the source code and task, weakening the reasoning capabilities required to perform a minimal change that remains executable.
Generally, we find that the particular prompt has less effect on the mutant status distribution than the model capability.

\subsection{Effectiveness of Test-Aware Prompting}
\begin{figure*}[thb]
\centering
\includegraphics[width=\linewidth]{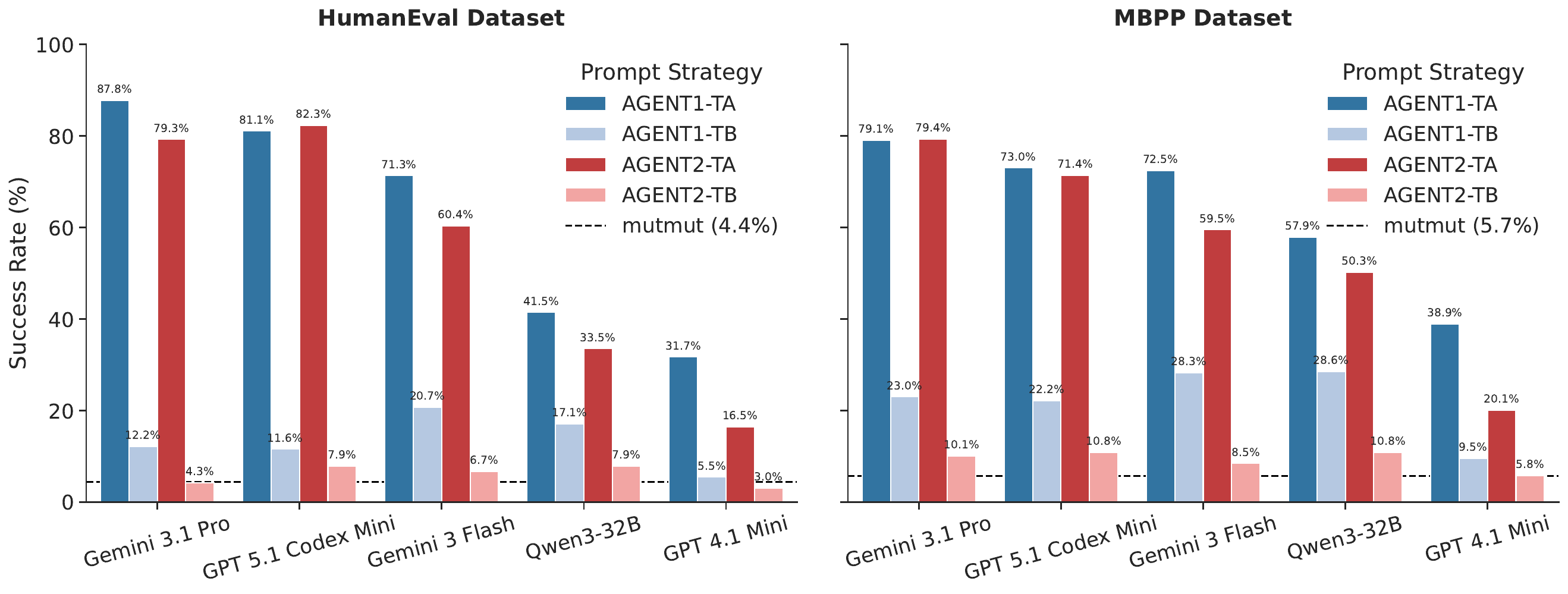}
\caption{Comparison of mutant success rates across models and prompting strategies. The dashed lines indicate the success rates of the traditional rule-based baseline (mutmut).}
\label{fig:rq1-effectiveness}
\end{figure*}

\Cref{fig:rq1-effectiveness} compares success rates across the five evaluated models. With mutmut achieving a success rate of only around 4.4\% on HumanEval and 5.7\% on MBPP (dashed line), this baseline is outperformed by all \glspl{llm}, whether prompted in test-aware or test-blind manner, with two exceptions: for \textsc{agent2-tb}, Gemini 3.1 Pro has a \successLabel rate of 4.3\% and GPT 4.1 Mini rate is 3.0\% on HumanEval. This suggests that even test-blind \glspl{llm} are capable of generating higher-value mutants than exhaustive mutmut generation, but test-awareness provides the greatest benefit.
In every model, the \textsc{ta} prompt variant outperforms \textsc{tb}. On HumanEval, \textsc{agent1-ta} increases the \successLabel rate for Gemini 3.1 Pro from 12.2\% to 87.8\% (7.2$\times$) and for GPT 5.1 Codex Mini from 11.6\% to 81.1\% (7.0$\times$). For \textsc{agent2-ta} as well, Gemini 3.1 Pro improves from 4.3\% (\textsc{tb}) to 79.3\% (\textsc{ta}) and GPT 5.1 Codex Mini improves from 7.9\% (\textsc{tb}) to 82.3\% (\textsc{ta}). Even for the weakest-performing model, GPT 4.1 Mini, \successLabel increases from 5.5\% to 31.7\% (\textsc{agent1}) and from 3.0\% to 16.5\% (\textsc{agent2}). On MBPP, the absolute gains are smaller but the trend remains. This consistency across two prompt families and benchmarks for all models shows that test-awareness alone is more crucial for improving mutant generation than the particular prompt phrasing.

Comparing across prompts, our results indicate that the free-form reasoning of \textsc{agent1} generally outperforms the structured template of \textsc{agent2}. The rigid constraint of \textsc{agent2-ta} to explicitly identify two to three blind spots appears to degrade the model's ability to produce base-test-compliant code, particularly for smaller models. For example, evaluating GPT 4.1 Mini across both prompt variants reveals that \textsc{agent2-ta} produces \rejectedLabel mutants on 156 tasks (48 in HumanEval, 108 in MBPP) where \textsc{agent1-ta} is able to evade the base tests for the same tasks (\successLabel or \undetectedLabel). This indicates that forcing explicit structural reasoning can hinder the adversarial search space in practice.

\subsection{Subtlety and Fault Quality}
\label{sec:subtelty-and-fault-quality}

\begin{figure*}[thb]
\centering
\includegraphics[width=\linewidth]{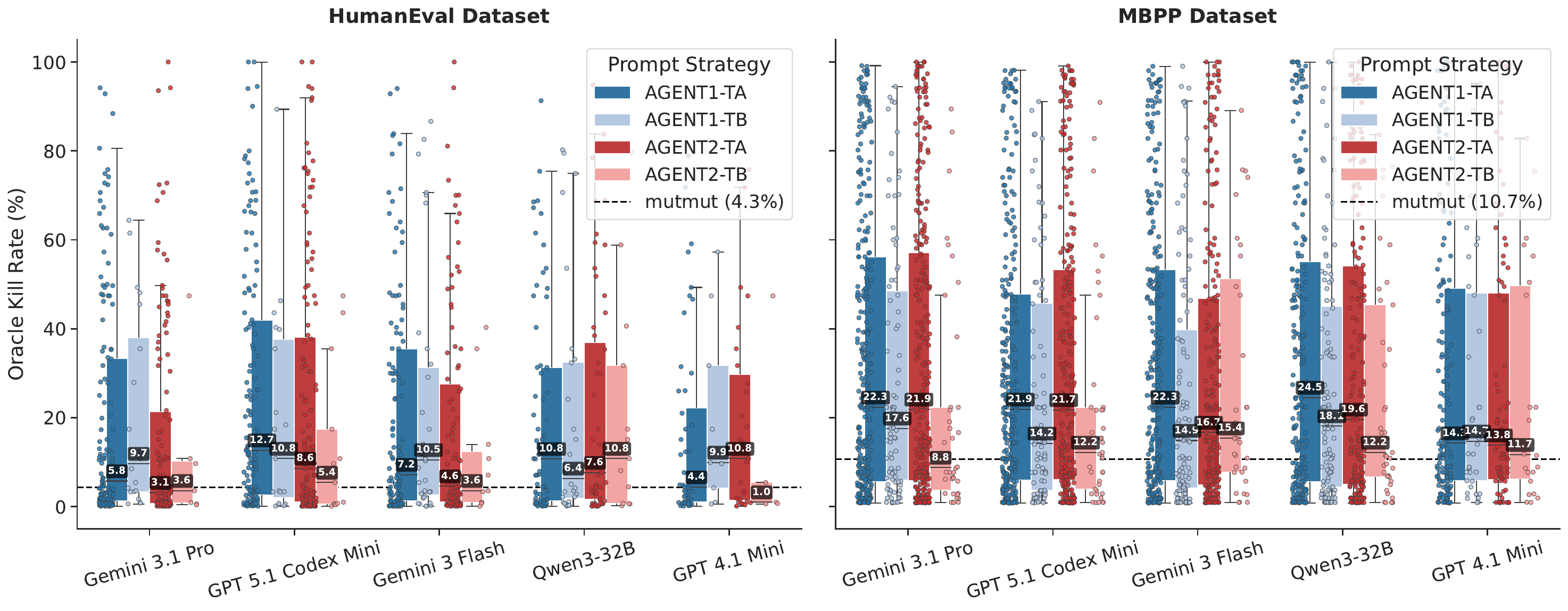}
\caption{Fault subtlety as measured by the oracle kill rate. A lower kill rate indicates subtler edge-case bugs that are more difficult for test suites to detect. The horizontal lines in each box and the small numbers above show the \emph{median} value. Thus, mutmut (dashed line) is also presented as the \emph{median} (not mean) in this plot. The whiskers represent 1.5$\times$ \gls{iqr}.}
\label{fig:rq2-subtlety}
\end{figure*}

\enlargethispage{\baselineskip}
\Cref{fig:rq2-subtlety} shows the distribution of oracle kill rates for the generated \successLabel mutants. We choose this visualization to identify whether certain settings produce more subtle bugs, indicated by a smaller number of extended test cases that catch the mutant. We observe that although test-aware models have a higher number of successful mutants, the median oracle kill rate for test-aware configurations falls between roughly 10-30\%, comparable to those produced by test-blind models. While the dense clustering of points in the lower quartile (\eg, see \textsc{Qwen3-32B}) indicates that test-aware prompting is highly capable of generating narrow, subtle edge-cases rather than catastrophic logical breakages, it is not clear from our experiments that \glspl{llm} produce significantly more or less subtle bugs than mutmut.

To identify the reason for the substantial number of \undetectedLabel mutants in \cref{fig:status-overview}, we manually review each such mutant (9 for HumanEval, 56 for MBPP) produced by the best-performing setting, Gemini 3.1 Pro with the \textsc{agent1-ta} prompt. We want to know whether undetected mutants represent failures of the generation pipeline (equivalent mutants) or failures of the extended test oracle (genuine but undetected faults). The classification scheme was derived deductively to cover all common mutation testing failures and successes in a non-overlapping manner. Due to resource constraints, the labeling process was conducted by a single author, and therefore inter-rater agreement was not measured. Nonetheless, this qualitative assessment provides valuable insight into the types of edge-case faults that can bypass even extensive automated oracles. Concretely, we establish a hierarchical classification scheme consisting of four equivalency classes and two non-equivalency classes (see \cref{tab:qualitative-analysis}), where equivalent mutants are programs that behave exactly as the original program on all possible inputs, representing a failed attempt to create a program fault.

The four distinctions among equivalent mutants are:
(1) Identical code or cosmetic changes such as removing comments or renaming variables,
(2) semantic refactoring, \ie, a change in algorithm procedure, but resulting in the same outputs,
(3) out-of-specification faults, for example a mutant that produces incorrect results on negative input values even though the task description explicitly restricts inputs to be positive, and
(4) dead code, where modifications only occur in unvisited execution branches.
Conversely, non-equivalent mutants introduce real faults that the extended oracle failed to catch, which we again divide into two categories:
(5) true oracle gaps, representing realistic bugs completely missed by the tests, such as HumanEval\#18 which asks how many times a given substring can be found in the original string, but never tests for an empty substring, and finally
(6) ambiguous faults, which are bugs that only manifest under poorly defined task specifications, such as MBPP\#87, which asks to merge three dictionaries without defining precedence for duplicate keys.
In summary, \undetectedLabel mutants of type (1)-(4) represent a failure of the mutant generator, (5) marks a failure in the extended test oracle, and (6) attributes the fault to the underspecified task description.

Out of the 65 \undetectedLabel mutants, we found that only 9 ($\sim$13.8\%) are equivalent mutants and a vast majority (49/65, $\sim$75.4\%) present serious edge-case bugs that slipped through the extended test cases. This shows that even EvalPlus -- the more robust version of HumanEval and MBPP -- does not provide guaranteed perfect test cases in these instances. The remaining 7 ($\sim$10.8\%) are classified as ambiguous faults, meaning they exploit underspecified edge cases where neither the task description nor the test suite clearly defines the expected behavior. This manual review confirms that rather than representing a failure of the mutation pipeline, the \undetectedLabel category highlights the \emph{strength} of test-aware \glspl{llm} by showing their ability to act as advanced adversarial testers that even expose flaws in the hidden test oracle, which is significantly stronger than the base tests.

\begin{table}[htpb]
\caption{Distribution of \undetectedLabel mutants across equivalent and non-equivalent classes. This qualitative analysis was manually performed on the results from Gemini 3.1 Pro with \textsc{agent1-ta}.}
\centering
\begin{tabularx}{\textwidth}{@{} l l X @{}}
\toprule
\textbf{Classification} & \textbf{Total} & \textbf{Instances} \\
\midrule
\multicolumn{3}{@{}l}{\emph{Equivalent Mutants}} \\
(1) Identical / cosmetic & 1 & HumanEval \#114 \\
(2) Semantic refactoring & 5 & MBPP \#\{103, 235, 463, 578, 765\} \\
(3) Out-of-spec fault      & 3 & HumanEval \#84, MBPP \#\{124, 445\} \\
(4) Dead code            & 0 & \\
\midrule
\multicolumn{3}{@{}l}{\emph{Non-Equivalent Mutants}} \\
(5) True oracle gap      & 49 & HumanEval \#\{16, 18, 36, 125, 132, 153, 162\}, \newline
MBPP \#\{9, 12, 58, 61, 77, 96, 123, 129, 130, 138, 139, 166, 168, 239, 245, 285, 294, 308, 410, 415, 421, 428, 479, 557, 566, 573, 577, 631, 635, 637, 730, 734, 737, 751, 755, 760, 772, 787, 793, 800, 804, 807\} \\
(6) Ambiguous fault      & 7 & MBPP \#\{87, 439, 580, 607, 733, 773, 799\} \\
\bottomrule
\end{tabularx}
\label{tab:qualitative-analysis}
\end{table}

\subsection{Cost Efficiency}
\begin{figure*}[thb]
\centering
\includegraphics[width=\linewidth]{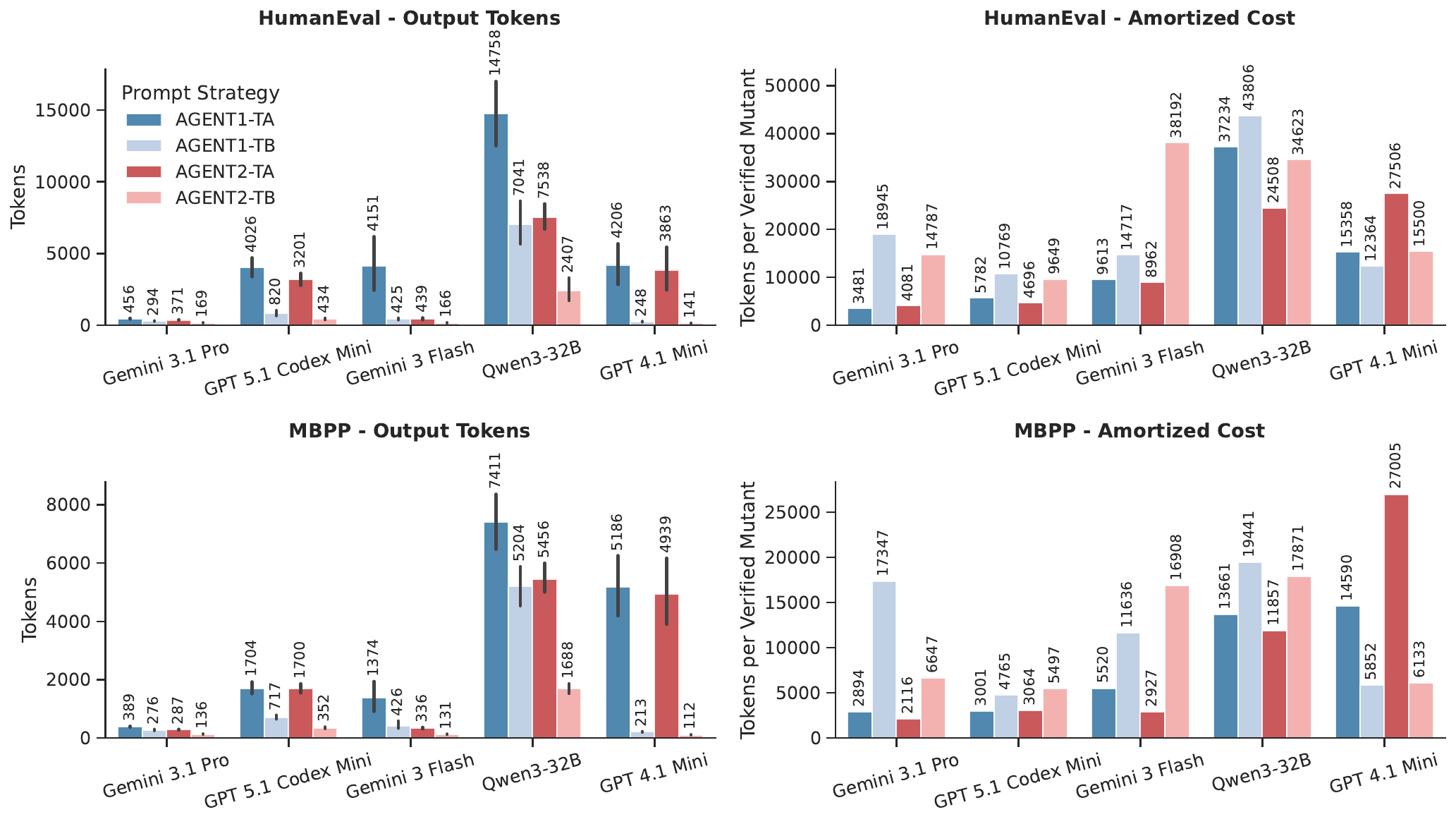}
\caption{Distribution of produced output tokens with 95\% confidence intervals (left) and amortized computational cost per successful mutant, representing the sum of total tokens (input + thinking + output) used for each mutant, divided by the number of \successLabel mutants (right).}
\label{fig:rq3-cost}
\end{figure*}

Naturally, \glspl{llm} are computationally more demanding than a rule-based mutant generation library like mutmut. We argue this price is worth paying for higher-quality results. Among the \gls{llm}-based approaches, we perform a fair cost comparison by considering the \emph{amortized} cost, \ie, the average total tokens consumed to produce a single \successLabel mutant. This analysis, shown in \cref{fig:rq3-cost}, reveals that test-aware prompting is more token-efficient per verified mutant than test-blind prompting for all models except GPT 4.1 Mini.

We also report the wall-clock time for running these experiments. For mutmut, the full HumanEval run took 6806.5\,s and generated 2841 mutants, which is around 2.4\,s per generated mutant and 54.0\,s per verified \successLabel mutant. On MBPP, mutmut took 8090.9\,s for 3427 mutants, around 2.4\,s per generated mutant and 41.9\,s per verified mutant. For GPT 5.1 Codex Mini with \textsc{agent1-ta}, the total runtime was 29.7\,s per problem on HumanEval and 16.4\,s on MBPP. This means that a single \gls{llm}-generated mutant is roughly 7--12$\times$ slower than a single mutmut mutant if we compare one generated candidate to one generated candidate. However, the runtime ranking reverses for cost per useful fault. Because test-aware prompting produces a larger proportion of verified faults, GPT 5.1 Codex Mini requires only 36.6\,s per verified mutant on HumanEval and 22.5\,s on MBPP, compared with 54.0\,s and 41.9\,s for mutmut. In other words, \gls{llm}-based mutation pays a higher per-candidate premium, but recovers that cost through much higher yield.
While \glspl{llm} incur higher upfront token and monetary costs, this is offset by the reduction in execution overhead. Executing large suites of trivial mutants is expensive in industrial pipelines; thus, investing compute into high-quality mutants allows for a substantially lower overall cost per verified fault.

\subsection{Iterative Prompting}
\label{sec:iterative-execution-results}

\begin{figure}[thb]
\centering
\includegraphics[width=\linewidth]{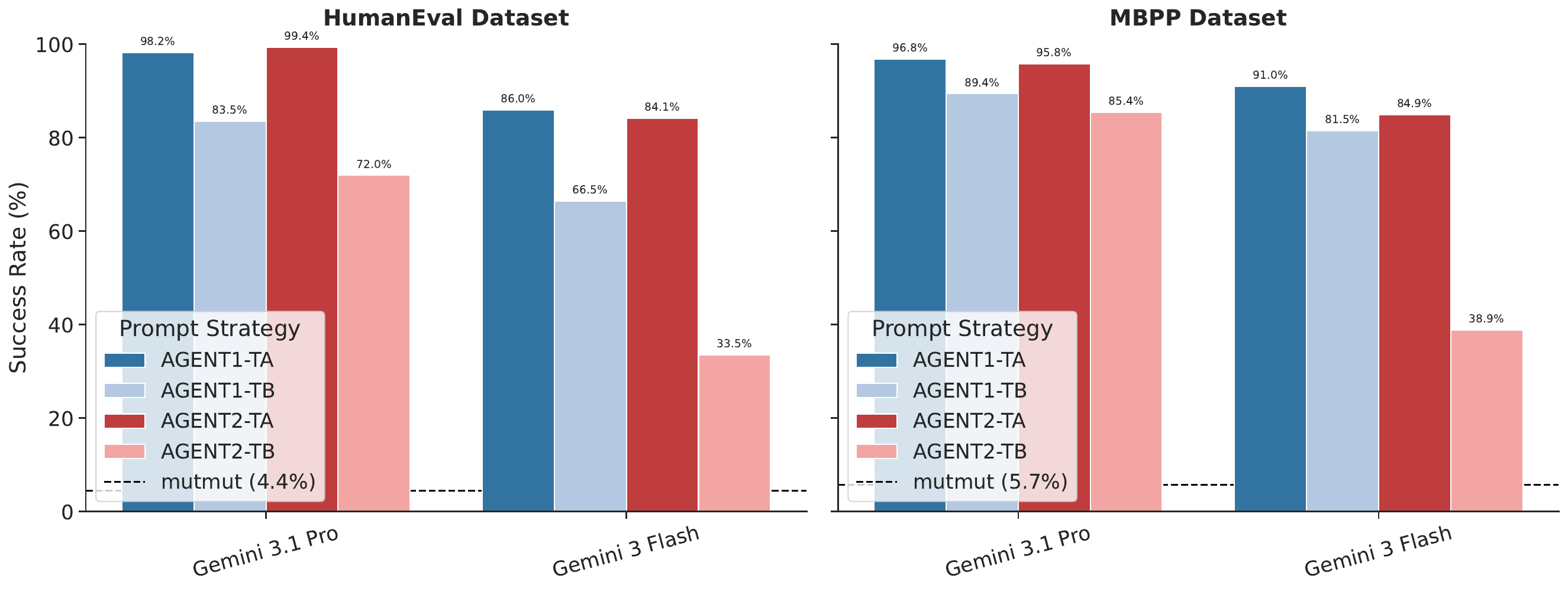}
\caption{Success rates under iterative prompting. The \textsc{ta}--\textsc{tb} gap narrows compared with the single-shot results in \cref{fig:rq1-effectiveness}, as repeated execution feedback partially compensates for missing test context (see \cref{tab:iterative-results}).}
\label{fig:rq1-effectiveness-iterative}
\end{figure}

\enlargethispage{-1\baselineskip}
While the single-shot results already show that test-awareness leads to subtle and effective mutant generation, a natural question is whether adding execution feedback can close the remaining gap. The answer is a clear yes: \cref{fig:rq1-effectiveness-iterative} shows results for an iterative variant of our pipeline in which the model follows a simple strategy to retry (up to 2 additional times) when a candidate fails the base tests: If a mutant is \invalidLabel, we show the execution error and ask for a syntax fix; if a mutant is \rejectedLabel, we include information on which base-test input failed by revealing the mutant output expected output; finally, if a mutant is \undetectedLabel, we encourage the model to take a different approach to avoid semantic equivalence to the original code.
Such iterative prompting narrows the \textsc{ta}--\textsc{tb} gap: Gemini 3.1 Pro \textsc{agent1-tb} rises from  12.2\% in single-shot to 83.5\%, approaching the iterative \textsc{ta} rate of 98.2\%. We see gains even on the already high-performing \textsc{ta} prompts (\eg, Gemini 3.1 Pro with \textsc{agent2-ta} on HumanEval reaches 163 out of 164 verified faults (99.4\%), an increase from 130 in the single-shot setting).
While the test-blind prompts use around twice the generation cost in the iterative setting, the test-aware iterative configurations require far fewer retries: Gemini 3.1 Pro averages 1.19--1.24 attempts per problem under \textsc{ta} versus 2.27--2.46 under \textsc{tb}. This reinforces the cost-efficiency advantage of test-awareness. These results suggest that single-shot test-aware prompting captures the majority of the achievable gain and iteration offers diminishing but still meaningful returns, particularly for converting the remaining 10--15\% of problems into verified faults.

\begin{table*}[thb]
\caption{Iterative prompting results for Gemini 3.1 Pro and Gemini 3 Flash on HumanEval and MBPP. Unlike the single-shot setting, the pipeline retries until the maximum attempt limit (in our case, 3) is reached. For each configuration, we report total generated mutants (Gen), counts of \invalidLabel (Inv), \rejectedLabel (Rej), \successLabel (Succ), and \undetectedLabel (Und) mutants, and the average number of attempts per problem (Att).}

\centering
\scriptsize
\setlength{\tabcolsep}{2.3pt}
\renewcommand{\arraystretch}{1.1}

\resizebox{\textwidth}{!}{
\begin{tabular}{llcccccccccccccccccccccccc}
\toprule

& &
\multicolumn{6}{c}{\textsc{agent1-ta}} &
\multicolumn{6}{c}{\textsc{agent2-ta}} &
\multicolumn{6}{c}{\textsc{agent1-tb}} &
\multicolumn{6}{c}{\textsc{agent2-tb}} \\

\cmidrule(lr){3-8}
\cmidrule(lr){9-14}
\cmidrule(lr){15-20}
\cmidrule(lr){21-26}

\textbf{Dataset} & \textbf{Model}
& Gen & Inv & Rej & Succ & Und & Att
& Gen & Inv & Rej & Succ & Und & Att
& Gen & Inv & Rej & Succ & Und & Att
& Gen & Inv & Rej & Succ & Und & Att \\

\midrule

\multirow{2}{*}{\textbf{HumanEval}}

& gemini-3.1-pro
& 195 & 0 & 0 & 161 & 3 & 1.19
& 203 & 0 & 0 & 163 & 1 & 1.24
& 373 & 0 & 9 & 137 & 18 & 2.27
& 404 & 1 & 18 & 118 & 27 & 2.46 \\

& gemini-3-flash
& 247 & 0 & 0 & 141 & 23 & 1.51
& 270 & 1 & 0 & 138 & 25 & 1.65
& 379 & 0 & 16 & 109 & 39 & 2.31
& 445 & 1 & 37 & 55 & 71 & 2.71 \\

\midrule

\multirow{2}{*}{\textbf{MBPP}}

& gemini-3.1-pro
& 469 & 0 & 0 & 366 & 12 & 1.24
& 490 & 0 & 0 & 362 & 16 & 1.30
& 735 & 0 & 4 & 338 & 36 & 1.94
& 846 & 0 & 5 & 323 & 50 & 2.24 \\

& gemini-3-flash
& 533 & 1 & 0 & 344 & 33 & 1.41
& 612 & 5 & 1 & 321 & 51 & 1.62
& 773 & 0 & 13 & 308 & 57 & 2.04
& 1011 & 3 & 46 & 147 & 182 & 2.67 \\

\bottomrule
\end{tabular}
}

\label{tab:iterative-results}
\end{table*}

\section{Discussion}
\label{sec:discussion}

Our experiments demonstrate that providing \pgls{llm} with test context pushes it to act as an adversarial agent that targets semantic blind spots rather than blindly generating equivalent faults. We further analyze these findings by examining the trade-offs between computational cost and human effort required to find a verified fault, why test-awareness works even in a single non-iterative prompt, how the results improve with an iterative feedback loop, and the limitations of deploying this approach in a real development environment.

\subparagraph*{Human Time \vs Compute Time Trade-off.}
Generating a few \quotes{smart} mutants is more valuable than a large number of low-quality (redundant, equivalent, trivial)  mutants. Although the computational cost in generating mutants using mutmut is far lower than \gls{llm}-based mutation testing, the human effort required to review these mutants often surpasses the computational efficiency. Petrovi\'{c} et al.~\cite{petrovic2018industrial} perform a cost-benefit analysis of mutation testing in industrial software development applications. They argue that the goal of mutation testing is to provide developers with \quotes{productive} mutants (that indicate genuine bugs) rather than achieving mutation adequacy by strengthening the test suite against all \quotes{unproductive} redundant mutants. This supports the intuition behind our \emph{test-aware} approach that a mutant generator becomes more useful when it produces fewer low-value candidates and more mutants that expose meaningful testing gaps. Our results fit this picture well. Test-aware \gls{llm} mutation does not win by being cheaper per generated candidate; it wins when cost is measured per useful, verified fault. In this regime, \quotes{smart} mutants become practically attractive because less human effort is required in the process.

\subparagraph*{Model Selection.}
The choice of \gls{llm} is an important practical consideration, because the effectiveness of our approach depends on it. For example, one test-aware instance (GPT 4.1 Mini \textsc{agent2-ta}) underperforms compared to the test-blind alternative (\textsc{agent1-tb} with Gemini 3 Flash and Qwen3-32B). Besides success rate, the fault subtlety and token cost are also affected. Our results suggest practical guidelines for deploying test-aware mutation: When the computational budget is strict, smaller models like GPT 5.1 Codex Mini save token costs. However, to maximize the discovery of subtle edge-case faults, frontier models like Gemini 3.1 Pro are recommended despite the higher API costs.

\subparagraph*{Why Single-Prompt Test-Awareness Helps.}
The most clear finding in this study is that simply providing the base tests in the prompt only produces more effective, subtle mutants over test-blind generation. This improvement is visible not only in overall success rates, but also in the status distribution: Moving from test-blind to test-aware prompting shifts a substantial portion of outputs from \rejectedLabel to \successLabel. The interpretation for such a trend is that once the model can see which behaviors are already checked, it becomes better at avoiding trivial mutations that are immediately killed by the base suite. In that sense, the main value of test-awareness is not only that it helps the model generate bugs, but also that it helps it generate bugs specifically aligned with existing test gaps.

\subparagraph*{The Dependency on Base Test Quality.}
Our results show that even few test cases can provide useful context for \gls{llm}-based mutant generation. In EvalPlus, where each problem has only few (7.7 tests on average per HumanEval problem and 3 for MBPP) base tests, access to those tests still substantially improves effectiveness over the test-blind setting. At the same time, this also highlights a limitation of the approach: \Pgls{llm} can only exploit shortcomings that are actually expressed in the provided tests. Because we did not systematically vary base-test quality or coverage, this study does not establish exactly how performance changes as the initial suite becomes stronger or weaker. That dependency remains an important direction for future work.

\subparagraph*{No Oracle in the Real World.}
Our evaluation relies on the extended EvalPlus tests as an oracle for determining whether a mutant that survives the base tests is actually non-equivalent. This is useful for controlled benchmarking, but it is not representative of most real-world development settings, where no such oracle is available. In practice, a surviving mutant would not automatically come with a ground-truth label; instead, it would signal that the current test suite may be missing a behavior worth checking. The main value of our setup is therefore methodological: It lets us verify that adding tests to the prompt helps \glspl{llm} find faults that the original suite misses. In real workflows, these mutants should be interpreted primarily as concrete suggestions for stronger tests and developer inspection rather than automatically validated defects. Identifying high-quality mutants makes it generally easy to specify an additional high-quality unit test to strengthen the test suite.

\subparagraph*{Practical Implications.}
Our findings indicate that test-aware \glspl{llm} can aid as review assistants in \gls{ci}, directly improving software engineering workflows. During pull requests, models could analyze modified code and tests to generate targeted mutants, providing developers actionable suggestions for stronger unit tests.

\section{Threats to Validity}

We identify and address the following potential threats to validity in our work:
\begin{alphaenumerate}
\item \textbf{Internal Validity: Sparse Base Tests.}
Bypassing only a handful of human-written base tests may appear to be a low bar for modern \glspl{llm}. However, we view this circumstance as a plausible representation of real-world codebases, where developers often have only basic assertions in place, before moving on to subsequent tasks. The aim of this study is not to show that \glspl{llm} are better than humans at generating mutants, but rather that they can achieve what a human developer with abundant time would do: uncover gaps in a test suite to make it stronger.
\item \textbf{Construct Validity: Lack of Human Review.}
Our evaluation pipeline is fully autonomous, which raises the question how reliable the mutant classification is. We highlight our careful experimental setup: The four classes are designed to be mutually exclusive and cover all distinct scenarios, with \invalidLabel, \rejectedLabel, and \successLabel being clearly defined and automatically verifiable. Only the last category leaves room for error, considering that \undetectedLabel mutants can either be fully valid code or contain a bug sneaky enough to go unnoticed by the strong extended test suite. To mitigate this threat, we perform a qualitative analysis, see \cref{sec:subtelty-and-fault-quality}. We find that most of such undetected cases in the best model-prompt combination are in fact serious faults, revealing that if we had a perfect oracle, this would only further improve the success rates. While EvalPlus serves as a robust automated oracle, we emphasize that it is a proxy rather than a definitive ground-truth measure of mutant quality in industrial settings.
\item \textbf{External Validity: Benchmark Limitations.} 
Our study evaluates self-contained Python benchmarks rather than complex production systems. We consider this sandbox setting a crucial first step, deliberately using EvalPlus as a controlled pipeline for reliable and transparent results. However, industrial codebases introduce technical complexities such as larger test suites, stateful functions, and multi-file dependencies. This could exceed \pgls{llm}'s context window and dilute its focus. Scaling to production-level datasets like Defects4J~\cite{just2014defects4j} may require iterative prompting to navigate stateful executions and \gls{rag}~\cite{lewis2020retrieval} to fetch relevant unit tests.
\item \textbf{External Validity: Model Constraints.}
\Gls{llm} performance inherently varies across models and provider-side policies. We mitigate this threat by evaluating five diverse \glspl{llm}, confirming that the benefits of test-aware prompting generally hold across different models. A discussion on the practical trade-offs of model selection is provided in \cref{sec:discussion}.
\end{alphaenumerate}

\section{Conclusion \& Future Work}

This paper explores whether a test-aware prompting strategy can turn \glspl{llm} into targeted mutant generators rather than generic bug injectors. Across HumanEval and MBPP, the answer is yes: Giving the model access to the existing tests substantially improves verified fault generation over both matched test-blind prompting and the rule-based baseline mutmut. Two broader findings stand out. First, the largest gain from test-awareness is not a universal reduction in oracle kill rate, but a significant increase in the number of verified, useful mutants produced at comparable subtlety. Second, although test-blind prompting can reduce tokens per generation due to less input and thinking tokens, test-aware prompting is more cost-effective per verified mutant because it wastes less computation on outputs that never become useful faults. Overall, our results suggest that test-aware prompting is a practical and lightweight alternative to both exhaustive rule-based mutation and heavier iterative multi-agent systems. More broadly, they indicate that when existing tests are available, \glspl{llm} can use them not only as a constraint, but also as a signal for where the test suite is still weak. This finding could substantially improve automated mutant generation, and ultimately lead to stronger test suites.

Our empirical findings pave the way for future work. For example, researchers with access to well-tested industrial code repositories could evaluate our pipeline by splitting their test suite into a handful of base tests and leaving the rest as oracle tests, pretending these are not yet part of the test suite. We hypothesize that this would replicate our finding that test-awareness leads to an increase in success rate.
Additionally, as \glspl{llm} scale, further research is needed to quantify the computational and environmental trade-offs of this approach. This includes comparing the energy consumption and CO$_2$ emissions of local versus cloud-hosted inference, and analyzing whether next-generation models with extended reasoning capabilities can further optimize the amortized cost per fault.

\section*{Data Availability Statement}

We make all code and data publicly available as open artifacts. Our code repository is available at \url{https://github.com/codingWhale13/test-aware-mutants}. The generated data is available at \url{https://zenodo.org/records/20195774} under a CC-BY 4.0 license. To reproduce the plots in this paper, rename the unzipped folder to \quotes{result}, place it at the root of the code repository, and follow the instructions in the repository's README file.

\bibliography{references}

\end{document}